# Evolution of Cloud Storage as Cloud Computing Infrastructure Service

R. Arokia Paul Rajan[1], S. Shanmugapriyaa[2]
[1](Department of Computer Science, Pope John Paul II College of Education, Pondicherry, India)
[2](Department of Computer Science, Pondicherry University, Pondicherry, India)

***ABSTRACT :*** *Enterprises are driving towards less cost, more availability, agility, managed risk - all of which is accelerated towards Cloud Computing. Cloud is not a particular product, but a way of delivering IT services that are consumable on demand, elastic to scale up and down as needed, and follow a pay-for-usage model. Out of the three common types of cloud computing service models, Infrastructure as a Service (IaaS) is a service model that provides servers, computing power, network bandwidth and Storage capacity, as a service to their subscribers. Cloud can relate to many things but without the fundamental storage pieces, which is provided as a service namely Cloud Storage, none of the other applications is possible. This paper introduces Cloud Storage, which covers the key technologies in cloud computing and Cloud Storage, management insights about cloud computing, different types of cloud services, driving forces of cloud computing and cloud storage, advantages and challenges of cloud storage and concludes by pinpointing few challenges to be addressed by the cloud storage providers.*

## I. INTRODUCTION

Cloud computing transforms the way in which current enterprises IT infrastructure is constituted and managed through consumable services such as infrastructure, platform, and applications. It will convert the IT infrastructure from a "factory" into a "supply chain" model. It is a type of computing that provides simple, on-demand access to pools of highly elastic computing resources [1]. These resources are provided as a service over a network, often the Internet. Cloud enables the consumers of the technology to think of computing as effectively limitless, of minimal cost, and reliable, as well as not to be concerned about how it is constructed, how it works, who operates it, or where it is located. The cloud computing model [4,5] is enabled by the ongoing standardization of underlying technologies like virtualization, Service Oriented Architecture (SOA), and Web 2.0.

Cloud computing is a style of computing where computing resources are easy to obtain and access, simple to use, cheap, and just work. Cloud is not a point product or a singular technology, but a way to deliver IT resources in a manner that provides self-service, on-demand and pay-per-use consumption. Utilizing cloud delivers time and cost savings. Cloud involves the subscriber and the provider. The service provider can be a company's internal IT group, a trusted third party or a combination of both. The subscriber is anyone who uses the services. By making data available in the cloud, it can be more easily and ubiquitously accessed, often at much lower cost, increasing its value by enabling opportunities for enhanced collaboration, integration, and analysis on a shared common platform.

## II. KEY TECHNOLOGIES

**2.1 Types of Cloud**

In a cloud computing system, there's a significant workload shift. Local computers no longer have to do all the heavy lifting when it comes to running applications. The network of computers that make up the cloud handles them instead which leads in reduction of hardware and software demands on the user's side. A typical cloud computing [6] architecture is given in Fig.1. The only thing the user's computer needs to be able to run is the cloud computing systems interface software, which can be as simple as a Web browser, and the cloud's network takes care of the rest.

There are three common types of clouds available namely, Private, Public and Hybrid cloud which is represented in Fig. 2. A private cloud is based upon a pool of shared resources, whose access is limited within organizational boundaries. The resources are accessed over a private and secured intranet, and are all owned and controlled by the company's IT organization. In essence, the cloud computing business model [7] is brought and managed in-house to enable shared IT services. A public cloud is a domain where the public Internet is used to obtain cloud services. The resources that make up those services are owned by the respective cloud service providers. Some examples include Salesforce.com, Google App Engine and Google search, Microsoft Azure, and Amazon Web services such as EC2

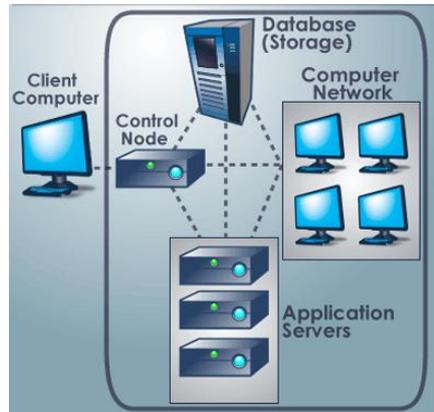

Fig. 1 A typical Cloud computing system

A Hybrid cloud is a combination of private and public clouds, where services from each domain are consumed in an integrated fashion and include an extended relationship with the selected external service providers.

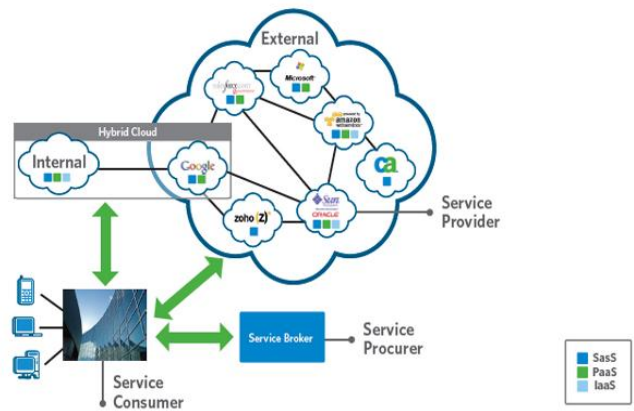

Fig. 2 Three types of Cloud computing model

**2.2 Cloud Computing Service Models**
Private and Public clouds serve as the backbone for a variety of different cloud computing service models given in Fig.3. Currently the industry has been successfully adopting three common types of cloud computing service models.
Infrastructure as a Service (IaaS), is a service model around servers (compute power), storage capacity, and network bandwidth. Examples include Amazon EC2 and S3, Rackspace, AT&T, and Verizon.
Platform-as-a-Service (PaaS) provides an externally managed platform for building and deploying applications and services. This model typically provides development tools such as databases and development studios for working with the supplied frameworks, as well as the infrastructure to host the built application. Examples include Force.com, Microsoft Azure, and Google App Engine.
Software-as-a-Service (SaaS) is simply having a software system running on a computer that doesn't belong to the customer and isn't on the customer's premises. It is based on the concept of renting an application from a service provider rather than buying, installing and running software yourself.

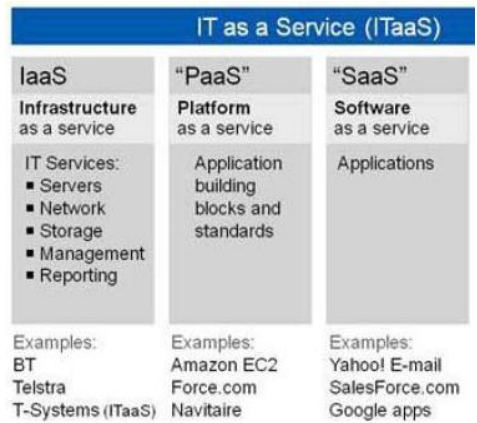

Fig. 3 Three types of Cloud service models

**2.3 Key benefits of Cloud Computing**

Management Insight, NH, USA, which is a dedicated market research consulting firm, conducted a study (6) on the impact of Cloud services in the market. This study was sponsored by CA Technologies, New York, USA. The statistical data (given in Fig.4 & 5) has revealed the following facts.

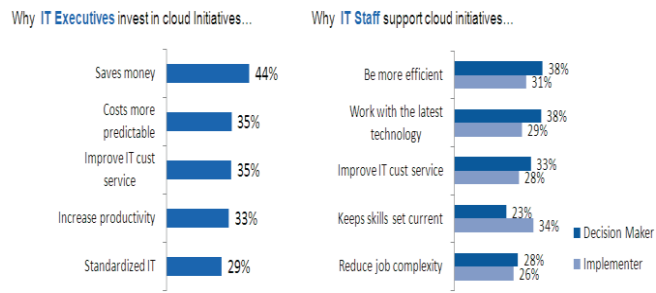

Fig. 4 IT personnel attitude towards the Cloud

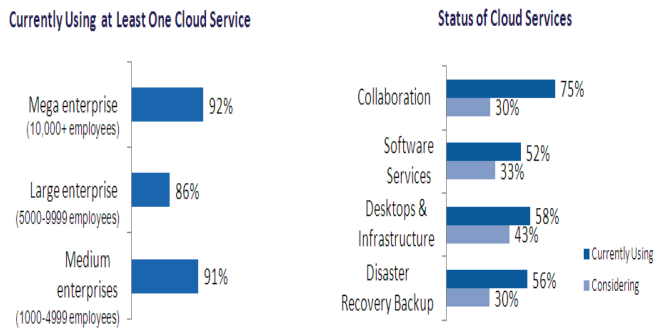

Fig. 5 Usage of Cloud services in the market

Cloud computing offers the following advantages to the enterprises:

- Lower costs: All resources, including expensive networking equipment, servers, IT personnel, etc. are shared, resulting in reduced costs, especially for small to mid-sized applications.
- Shifting Capital Expenses to Operational Expenses: Cloud computing enables companies to shift money from capital expenses to operating expenses, which ultimately allows the enterprise to focus their money and resources on innovation.
- Agility: Provisioning on-demand enables faster setup on an as-needed basis. When a project is funded, customer can initiate service, and then if the project is over, they can simply terminate the cloud contract.
- Scalability: Many cloud services can smoothly and efficiently scale to handle the growing nature of the business with a more cost effective pay-as-you-go model. This is also known as elasticity.

- Simplified maintenance: Patches and upgrades are rapidly deployed across the shared infrastructure, as well as the backups.
- Diverse platform support: Many cloud computing services offer built-in support for a rich collection of client platforms including browsers, mobile, and more. This diverse platform support enables applications to reach a broader category of users.
- Faster development: Cloud computing platforms provide many of the core services that, under traditional development models, would normally be built in house. These services, plus templates and other tools can significantly accelerate the development cycle.
- Large scale prototyping / Testing: Cloud computing makes large scale prototyping and load testing much easier. A client can easily spawn 1,000 servers in the cloud to load test your application and then release them as soon as they are done, and then try doing that with owned or corporate servers.

### III. CLOUD STORAGE

Rapid data growth and the need to keep it safer and longer will require organizations to integrate how they manage and use their data, from creation to end of life. Now there is an opportunity to store all our data in the internet. Those off-site storages are provided and maintained by the third parties through the Internet which is represented in Fig. 6. Cloud storage offers a large pool of storage was available for use, with three significant attributes: access via Web services APIs on a non persistent network connection, immediate availability of very large quantities of storage, and pay for what you use. It supports rapid scalability [2].

**3.1 Evolution of Cloud Storage**

Cloud storage is an offering of cloud computing. Fig. 7 shows the evolution of Cloud Storage based on traditional network storage and hosted storage. Benefit of cloud storage is the access of your data from anywhere. Cloud storage providers provide storage varying from small amount of data to even the entire warehouse of an organization. Subscriber can pay to the cloud storage provider for what they are using and how much they are transferring to the cloud storage.

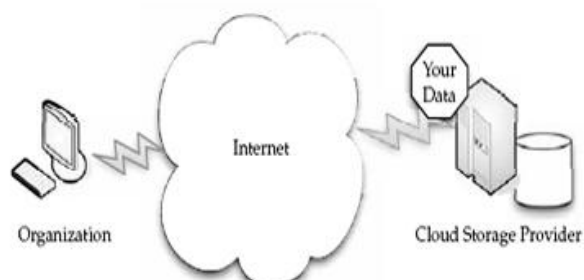

Fig. 6 Simple cloud storage model

Basically the cloud storage subscriber copies the data into any one of the data server of the cloud storage provider. That copy of data will be made available to all the other data servers of the cloud storage provider featuring redundancy in the availability which ensures that the data of the subscriber is safe even anything goes wrong. Most systems store the same data on servers that use different power supplies.

**3.2 Benefits of Cloud storage:**
- No need to invest any capital on storage devices.
- No need for technical expert to maintain the storage, backup, replication and importantly disaster management.
- Allowing others to access your data will result with collaborative working style instead of individual work.

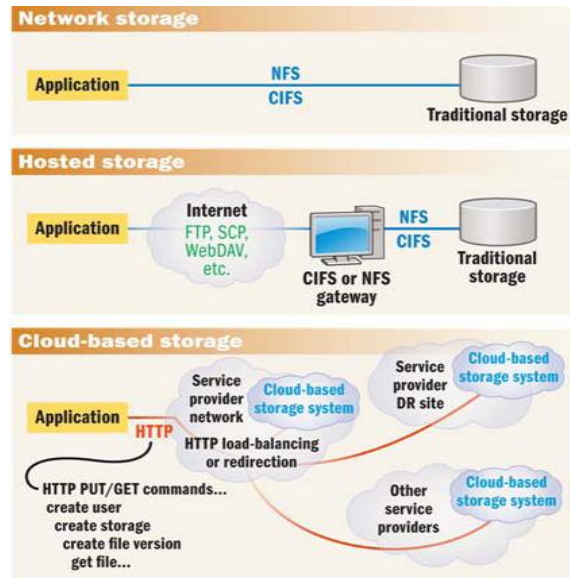

Fig. 7 Evolution of Cloud Storage

**3.3 Cloud Storage Reference Model**

    The appeal of cloud storage is due to some of the same attributes that define other cloud services: pay as you go, the illusion of infinite capacity (elasticity), and the simplicity of use/management [7]. It is therefore important that any interface for cloud storage support these attributes, while allowing for a multitude of business cases and offerings, long into the future.

    The model created and published by the Storage Networking Industry Association (SNIA) [7,8] shows multiple types of cloud data storage interfaces are able to support both legacy and new applications. All of the interfaces allow storage to be provided on demand, drawn from a pool of resources. The capacity is drawn from a pool of storage capacity provided by storage services. The data services are applied to individual data elements as determined by the data system metadata. Metadata specifies the data requirements on the basis of individual data elements or on groups of data elements (containers).

    As shown in Fig. 8, Cloud Data Management Interface (CDMI) is the functional interface that applications will use to create, retrieve, update and delete data elements from the cloud. As part of this interface the client will be able to discover the capabilities of the cloud storage offering and use this interface to manage containers and the data that is placed in them. In addition, metadata can be set on containers and their contained data elements through this interface. It is expected that the interface will be able to be implemented by the majority of existing cloud storage offerings today. This can be done with an adapter to their existing proprietary interface, or by implementing the interface directly. In addition, existing client libraries such as eXtensible Access Method (XAM) can be adapted to this interface as show in Fig. 8.

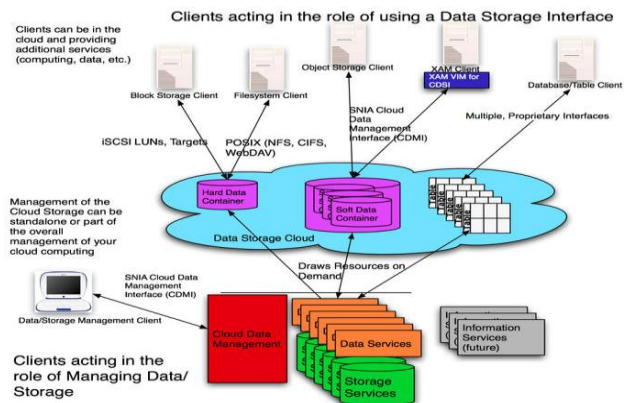

Fig. 8 Cloud Storage Reference model

This interface is also used by administrative and management applications to manage containers, accounts, security access, monitoring/billing information and even for storage that is accessible by other protocols. The capabilities of the underlying storage and data services are exposed so that clients can understand the offering. Conformant cloud offerings may offer a subset of either interface as long as they expose the limitations in the capabilities part of the interface.

**3.4 Cloud Storage API**
A Cloud Storage Application Programming Interface (API) is a method for access to and utilization of a cloud storage system [3,11]. The most common of these kinds are REST (REpresentational State Transfer) although there are others, which are based on SOAP (Simple Object Access Protocol). All these APIs are associated with establishing requests for service via the Internet. REST is a concept widely recognized as an approach to "quality" scalable API design.

One of the most important features of REST is that it is a "stateless" architecture [10, 11]. This means that everything needed to complete the request to the storage cloud is contained in the request, so that a session between the requestor and the storage cloud is not required. It is very important because the Internet is highly latent (it has an unpredictable response time and it is generally not fast when compared to a local area network). REST is an approach that has very high affinity to the way the Internet works. Traditional file storage access methods that use NFS (network files system) or CIFS (Common Internet File System) do not work over the Internet, because of latency.

Cloud Storage is for files, which, some refer to as objects, and others call unstructured data. Think about the files stored on your PC, like pictures, spreadsheets and documents. These have an extraordinary variability, thus unstructured. The other kind of data is block or structured data. Think data base data, data that feeds transactional system that require a certain guaranteed or low-latency performance. Cloud Storage is not for this use case. Industrial Design Centre (IDC) estimates that approximately 70% of the machine stored data in the world is unstructured, and this is also the fastest growing data type. So, Cloud Storage is storage for files that is easily accessed via the Internet. This does not mean you cannot access Cloud Storage on a private network or LAN, which may also provide access to a storage cloud by other approaches, like NFS or CIFS. It does mean that the primary and preferred access is by a REST API.

REST APIs are language neutral and therefore can be leveraged very easily by developers using any development language they choose. Resources within the system may be acted on through a URL. So, an API is not a "programming language", but it is the way a programming language is used to access a storage cloud. REST APIs are also about changing the state of resource through representations of those resources. They are not about calling web service methods in a functional sense. The key differences between different Cloud Storage APIs are the URLs defining the resources and the format of the representations. Amazon S3 APIs, Eucalyptus APIs, Rackspace Cloud Files APIs, Mezeo APIs, Nivanix APIs, Simple Cloud API, along with the standards proposed by the Storage Networking Industry Association (SNIA) Cloud Storage Technical Work Group, and more.

## IV. ISSUES IN CLOUD STORAGE

Cloud storage gets the attention of IT managers with its comparatively low cost and ability to easily adjust capacity. Although cloud storage offers reduction in the capital investment cost, customers has to face some of the technical, integration, security and organizational issues at various levels. Also Fig. 9 shows how IT people are hesitating to take up the cloud services.

- Control over the Data: Since the data is residing outside the enterprise's infrastructure, it is perceived that the enterprise may loss the control over data. Although the concerns are largely hypothetical and psychological rather than actual, due to the immaturity of cloud services, standards on the delivery of services and their evolving business model, users may have genuine concerns about the service provider's viability and operational processes.
- Interoperability & Control: The complexity of using cloud storage is something many customers underestimate "It's not plug-and-play." Each vendor has different access methods, nonstandard APIs that make integrating applications, such as archiving or file shares with cloud storage, difficult and costly. Some vendors provide software clients that implement common network file sharing protocols such as Network File System (NFS) or Common Internet File System (CIFS) [11], but these are proprietary and cannot bridge between different cloud services. The lack of standard protocols for accessing cloud storage means there is no interoperability between cloud storage providers, greatly complicating the data migration.

- Performance & Security: Access to cloud data is obviously limited by network throughput and latency, and despite of drastic improvements in Internet performance, it is still poor in comparison to local network storage. Although some vendors attempt to enhance throughput with various local caching and compression techniques, these don't improve Internet latency. Data security is the biggest issue with cloud storage. If any possibility leakage, both in transfer and within a shared infrastructure, experts agree that using encryption on all data stored in a cloud is essential although depending on the application, this is easier said than done.
- Suitability of applications: A kind of more static data, inactive data, such as applications that include online backup and archiving is the best fit for cloud storage. The archiving kind of data works well in the cloud because the data changes less frequently. These data don't require high speed transactional access. Bulk data can be easily compressed using data reduction technologies as well as it can be easily encrypted. Applications like content delivery [2] of rich media such as video, audio, or image files, Web 2.0 applications, user files and email repositories are some examples for best fit into the cloud storage. The applications with low I/O performance and tolerance for low downtime are suitable residents in cloud storage.

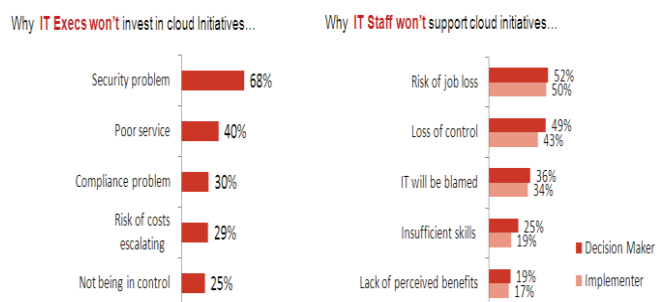

Fig. 9 Reasons why IT personnel hesitate to take up cloud services

## V. CONCLUSION

Cloud computing represents the next evolutionary step toward elastic IT. Cloud emergence transforms the way in which IT infrastructure is constituted and managed through consumable services for infrastructure, platform, and applications. This idea converts IT infrastructure from a "factory" into a "supply chain". There may be a stage to come when Internet is going to be the communication channel for mass media, then we cannot imagine a world without cloud storage because keeping ownership and maintaining huge volume of data on our own infrastructure is unimaginable. So automatically cloud storage will capture the entire market as we see rental houses leased to the tenants which is an unavoidable and a must situation in a growing populated city.

Cloud storage strategies and service models are still in its early stages. Standardization of service provider's service levels, pricing plans, data access methods, operational and security processes, emergency plans for data migration if the enterprise sooner or later wish to change vendors, improving the performance by opting better load balancing methodology are some of the thrust areas where future works on cloud storage can be focused.

## ACKNOWLEDGEMENT

The authors would like to acknowledge Dr.F.Sagayaraj Francis, Associate Professor, Department of Computer Science & Engineering, Pondicherry Engineering College, for his valuable support and encouragement.

# REFERENCES

### Journal Papers:
[1] Daniel J. Abadi, Data Management in the Cloud: Limitations and Opportunities, *IEEE Data Engineering Bulletin, Volume 32,* March 2009, 3-12.

[2] James Broberg, Rajkumar Buyya, Zahir Tari, MetaCDN: Harnessing 'Storage Clouds' for high performance content delivery, *Journal of Network and Computer Applications*, 1012–1022, 2009.

### Books:
[3] Eric A. Marks, Bob Lozano, *Executive's Guide to Cloud Computing* (John Wiley & Sons Inc, pp. 25-40, 2010).

[4] Anthony T. Velte, Toby J. Velte, Robert Elsenpeter, *Cloud Computing: A Practical Approach* (McGraw Hill Publications, pp 135 – 144, 2010).

[5] Linda Xu, Miklos Sandorfi and Tanya Loughlin, *Cloud Storage for Dummies* (Wiley Publishing, pp. 5-24, 2010).

[6] Wu Jiyi,Ping Lingdi,Pan Xuezeng.Cloud Computing: Concept and Platform,Telecommunications Science,12:23-30, 2009.

[7] Storage Networking Industry Association.Cloud Storage Reference Model,Jun.2009.

### Thesis:
[8] Srikumar Venugopal, *Scheduling Distributed Data-Intensive Applications on Global Grids*, Doctoral diss., Department of Computer Science and Software Engineering, The University of Melbourne, Australia, July 2006.

### Proceedings Papers:
[9] Nicolas Bonvin, Thanasis G. Papaioannou and Karl Aberer, A Self-organized, Fault-tolerant and Scalable replication scheme for Cloud storage, *SoCC '10 Proceedings of the 1st ACM symposium on Cloud computing,* New York, USA, 2010, 205-216.

[10] Wenying Zeng, Yuelong Zhao, Kairi Ou, Wei Song, Research on cloud storage architecture and key technologies, *ICIS '09: Proceedings of the 2nd International Conference on Interaction Sciences: Information Technology, Culture and Human*, ACM New York, NY, USA, 2009, 1044-1048.

[11] Luis M.Vaquero,Luis Rodero-Merino, Juan Caceres,Maik Lindner, A Break in the Clouds: Toward a Cloud Definition. *ACM SIGCOMM Computer Communication Review*, 2009, 50-55.

### Online:
[12] Lee Black, Jack Mandelbaum, Indira Grover, Yousuf Marvi, The arrival of "Cloud Thinking": How and Why Cloud Computing Has Come of Age in Large Enterprises, Management Insight Technologies. Online Available at: http://www.ca.com/~/media/files/whitepapers/the_arrival_of_cloud_thinking.aspx

[13] Deepak Kanwar, Cloud Storage: A Business Model for Service Providers - Should you enter the Cloud Storage Market?, Mezeo Software Inc. Online Available at: www.cloudstoragestrategy.com/bmodel

[14] Jonathan Hoppe, A Closer Look: Enterprise-Grade Cloud Storage. Online Available at: http://www.datacenterknowledge.com/archives/2011/03/17/a-closer-look-enterprise-grade-cloud-storage/

[15] Steve Lesem, Cloud Storage Strategy: Insights, Observations, and Next Practices on Cloud Storage and Services, November 2009. Online Available at: http://www.cloudstoragestrategy.com

[16] Mike Hogan, Cloud Computing & Databases: How databases can meet the demands of cloud computing, ScaleDB Inc., November 14, 2008 Online Available at: www.scaledb.com/pdfs/CloudComputingDaaS.pdf

[17] http://computer.howstuffworks.com/cloud-storage.htm

[18] http://cloudstoragestrategy.com/2009/11/understanding-cloud-storage-apis.html

[19] Intelligent Data Management, Dell Storage Product Group, IBM. Online Available at: http://i.dell.com/sites/content/business/smb/sb360/en Documents/wp-idm.pdf